\documentclass[preprint,aps,floats,subeqn,showpacs,amssymb]{revtex4}
\usepackage{epsfig}

\newcommand{\be}{\begin{equation}}
\newcommand{\ee}{\end{equation}}
\newcommand{\bea}{\begin{eqnarray}}
\newcommand{\eea}{\end{eqnarray}}
\newcommand{\bml}{\begin{mathletters}}
\newcommand{\eml}{\end{mathletters}}

\begin{document}

\tighten

\preprint{IUB-TH-0411}
\draft




\title{Non-abelian black strings}
\renewcommand{\thefootnote}{\fnsymbol{footnote}}
\author{Betti Hartmann\footnote{b.hartmann@iu-bremen.de}}
\affiliation{School of Engineering and Sciences, International University Bremen (IUB),
28725 Bremen, Germany}

\date{\today}
\setlength{\footnotesep}{0.5\footnotesep}

\begin{abstract}
Non-abelian black strings in a 5-dimensional Einstein-Yang-Mills
model are considered. 
The solutions are spherically symmetric non-abelian black holes
in 4 dimensions extended into an extra dimension and thus
possess horizon topology $S^2 \times \mathbb{R}$.
We find that several branches of solutions exist. In addition, we determine
the domain of existence of the non-abelian black strings.  
\end{abstract}

\maketitle

\section{Introduction}
The idea that we live in more than four dimensions 
stems from models of unification of all known forces.
In Kaluza-Klein theories \cite{kaluza,klein} 4-dimensional electromagnetism
and gravity are unified within a 5-dimensional gravity theory.
The 5th dimension is compactified on a circle of Planck length.
Similar ideas appear in (super)string theories, which are only consistent
in 10 and 26 dimensions in the case of superstring, respectively bosonic
string theory \cite{pol}.
Recently, so-called brane world models \cite{brane} have gained a lot of
interest. These assume the standard model fields to be
confined on a 3-brane embedded in a higher dimensional manifold.

A large number of higher dimensional black holes has been studied in
recent years. The first solutions that have been constructed
are the hyperspherical generalisations of well-known black holes
solutions such as the Schwarzschild and Reissner-Nordstr\"om solutions
in more than four dimensions \cite{tan} as well
as the higher dimensional Kerr solutions \cite{mp}.
In $d$ dimensions, these solutions have horizon topology $S^{d-2}$.

However, in contrast to 4 dimensions black holes with different horizon
topologies should be possible in higher dimensions. 
An example is a 4-dimensional
Schwarzschild black hole extended into one extra dimension, a so-called
Schwarzschild black string. These solutions have been discussed extensively
especially with view to their stability \cite{gl}.
A second example, which is important due to its implications for
uniqueness conjectures for black holes in higher dimensions, is the black ring
solution in 5 dimensions with horizon topology $S^2\times S^1$ \cite{er}.
 
The by far largest number
of higher dimensional
black hole solutions constructed so far are solutions of the
vacuum Einstein equations, respectively Einstein-Maxwell equations.
The first example of black hole solutions containing non-abelian
gauge fields have been discussed in \cite{bcht}. These are non-abelian
black holes solutions of a generalised 5-dimensional Einstein-Yang-Mills
system with horizon topology $S^3$.

Here we construct black hole solutions of an Einstein-Yang-Mills
model discussed previously in \cite{volkov,bh1,bch}. These solutions
are non-abelian black hole solutions in 4 dimensions extended
into one extra codimension.
We give the model including the ansatz, equations of motion and
boundary conditions in Section II. The numerical results are discussed
in Section III. The summary is given in Section IV.

\section{The Model}

The Einstein-Yang-Mills Lagrangian
in $d=(4+1)$ dimensions is
given by:

\begin{equation}
\label{action}
  S = \int \Biggl(
    \frac{1}{16 \pi G_{5}} R   - \frac{1}{4 e^2}F^a_{M N}F^{a M N}
  \Biggr) \sqrt{g^{(5)}} d^{5} x
\end{equation}
with the SU(2) Yang-Mills field strengths
$F^a_{M N} = \partial_M A^a_N -
 \partial_N A^a_M + \epsilon_{a b c}  A^b_M A^c_N$
, the gauge index
 $a=1,2,3$  and the space-time index
 $M=0,...,5$. $G_{5}$ and $e$ denote
respectively the $5$-dimensional Newton's constant and the coupling
constant of the gauge field theory. $G_{5}$ is related to the Planck mass
$M_{pl}$ by $G_{5}=M_{pl}^{-3}$ and $e^2$ has the dimension of 
$[{\rm length}]$.

If both the metric and matter fields
are independent on the extra coordinate $y$, the fields can be
parametrized as follows \cite{volkov}:
\begin{equation}
g^{(5)}_{MN}dx^M dx^N = 
e^{-\xi}g^{(4)}_{\mu\nu}dx^{\mu}dx^{\nu}
+e^{2\xi} dy^2 
\ , \ \mu , \nu=0, 1, 2, 3 
\end{equation}
and
\begin{equation}
\label{gauge}
A_M^{a}dx^M=A_{\mu}^a dx^{\mu}+ \Phi^a dy   \  . \
\end{equation}
$g^{(4)}$ is the $4$-dimensional metric tensor.

\subsection{The Ansatz}
Our aim is to construct non-abelian black holes, which are
spherically symmetric in 4 dimensions and are extended
into one extra dimension. The topology of these non-abelian
black strings will thus be $S^2\times \mathbb{R}$. Note that
we could also construct non-abelian black holes with
topology $S^2\times S^1$ if we would make the extra coordinate
$y$ periodic.
 
For the metric the Ansatz reads:

\begin{equation}
g^{(5)}_{MN}dx^M dx^N = 
e^{-\xi}\left[-A^{2}(r)N(r)dt^2+N^{-1}(r)dr^2+r^2 d\theta^2+r^2\sin^2\theta
d^2\varphi\right]
+e^{2\xi} dy^2 
\label{metric}
\ , \end{equation}
with
\begin{equation}
N(r)=1-\frac{2m(r)}{r} \ \ \ {\rm and} \ \ \ \xi=\xi(r) 
\ . \end{equation}
In these coordinates, $m(\infty)$ denotes the (dimensionful) mass per unit length
of the field configuration.\

For the gauge fields, we use the purely magnetic hedgehog ansatz
\cite{thooft} :
\begin{equation}
\label{ansatz1}
{A_r}^a={A_t}^a=0
\ , \end{equation}
\begin{equation}
\label{ansatz2}
{A_{\theta}}^a= (1-K(r)) {e_{\varphi}}^a
\ , \ \ \ \
{A_{\varphi}}^a=- (1-K(r))\sin\theta {e_{\theta}}^a
\ , \end{equation}
\begin{equation}
\label{higgsansatz}
{\Phi}^a=v H(r) {e_r}^a \ \ , 
\end{equation}
where $v$ is a mass scale.

\subsection{Equations of motion}
With the following rescalings:
\begin{equation}
x=evr \ \ , \ \ \mu=evm
\end{equation}
the resulting set of ordinary differential equations only depends
on the fundamental coupling $\alpha=4\pi \sqrt{G_{5}} v$.

The Einstein equations for the metric functions $N$, $A$ and $\xi$ then read:

\begin{equation}
\mu ' = \alpha^2 \left(e^{\xi}N(K')^2 + \frac{1}{2}N x^2(H')^2 
e^{-2\xi}+
\frac{1}{2x^2}(K^2-1)^{2} e^{\xi}+K^2 H^2 e^{-2\xi}\right)
+  \frac{3}{8}Nx^{2}(\xi ')^2  \ , 
\end{equation}
\begin{equation}
A'=\alpha^2 x A \left(\frac{2(K')^2}{x^2}e^{\xi}+
e^{-2\xi}(H')^2\right)+\frac{3}{4} x A(\xi ')^2
\ , \label{dgl5} \end{equation} 
\begin{eqnarray}
(x^2 AN\xi')' &=& \frac{4}{3}\alpha^2 
A \left[e^{\xi}\left(N(K')^2+\frac{(K^2-1)^2}{2 
x^2}\right)- 2 
e^{-2\xi}\left(\frac{1}{2}
N (H')^2 x^2+H^2 K^2\right) \right] 
\ , \label{dgl3} \end{eqnarray}
while the Euler-Lagrange equations for the matter functions $K$ and $H$ are given by:
\begin{equation}
(e^{\xi}ANK')'=A\left(e^{\xi}\frac{K(K^2-1)}{x^2}+e^{-2\xi}H^2 
K\right)
\ , \label{dgl1} \end{equation}
\begin{equation}
(e^{-2\xi}x^2 ANH')'=2e^{-2\xi}K^2 AH
\ , \label{dgl2} 
\end{equation}

where the prime denotes the derivative with respect to $x$.

\subsection{Boundary conditions}
The boundary conditions at the regular horizon $x=x_h$ read:
\begin{equation}
N(x_h)=0 \Rightarrow  \mu(x_h)=\frac{x_h}{2} 
\end{equation}
and
\begin{equation}
(N' K')|_{x=x_h}=\left[\frac{K(K^2-1)}{x^2}+e^{-3\xi} H^2 K\right]\vert_{x=x_h} \ ,
\end{equation}
\begin{equation}
(N' H')|_{x=x_h}=\left(\frac{2}{x^2}K^2 H\right)\vert_{x=x_h} \ ,
\end{equation}
\begin{equation}
(N' \xi')|_{x=x_h}=\left\{\frac{4}{3x^2}\alpha^2\left[e^{\xi}\left(N(K')^2+\frac{(K^2-1)^2}{2 
x^2}\right)- 2 
e^{-2\xi}\left(\frac{1}{2}
N (H')^2 x^2+H^2 K^2\right) \right] \right\} \vert_{x=x_h}  
\end{equation}
with $A(x_h) < \infty$.

At infinity, finiteness of the energy and asymptotic
flatness requires:
\begin{equation}
A(\infty)=1 \ \ , \ \ K(\infty)=0 \ \ , \ \ H(\infty)=1 \ \ , 
\ \ \xi(\infty)=0 \ .
\end{equation}

\section{Numerical results}
The set of equations has been solved numerically \cite{foot1}.
These black hole solutions can be interpreted as 4-dimensional
black holes sitting in the center of particle-like solutions
extended into one extra dimension. Due to the non-triviality
of the matter fields outside the regular horizon, these solutions
possess ``hair''.

\begin{figure}
\centering
\epsfysize=10cm
\mbox{\epsffile{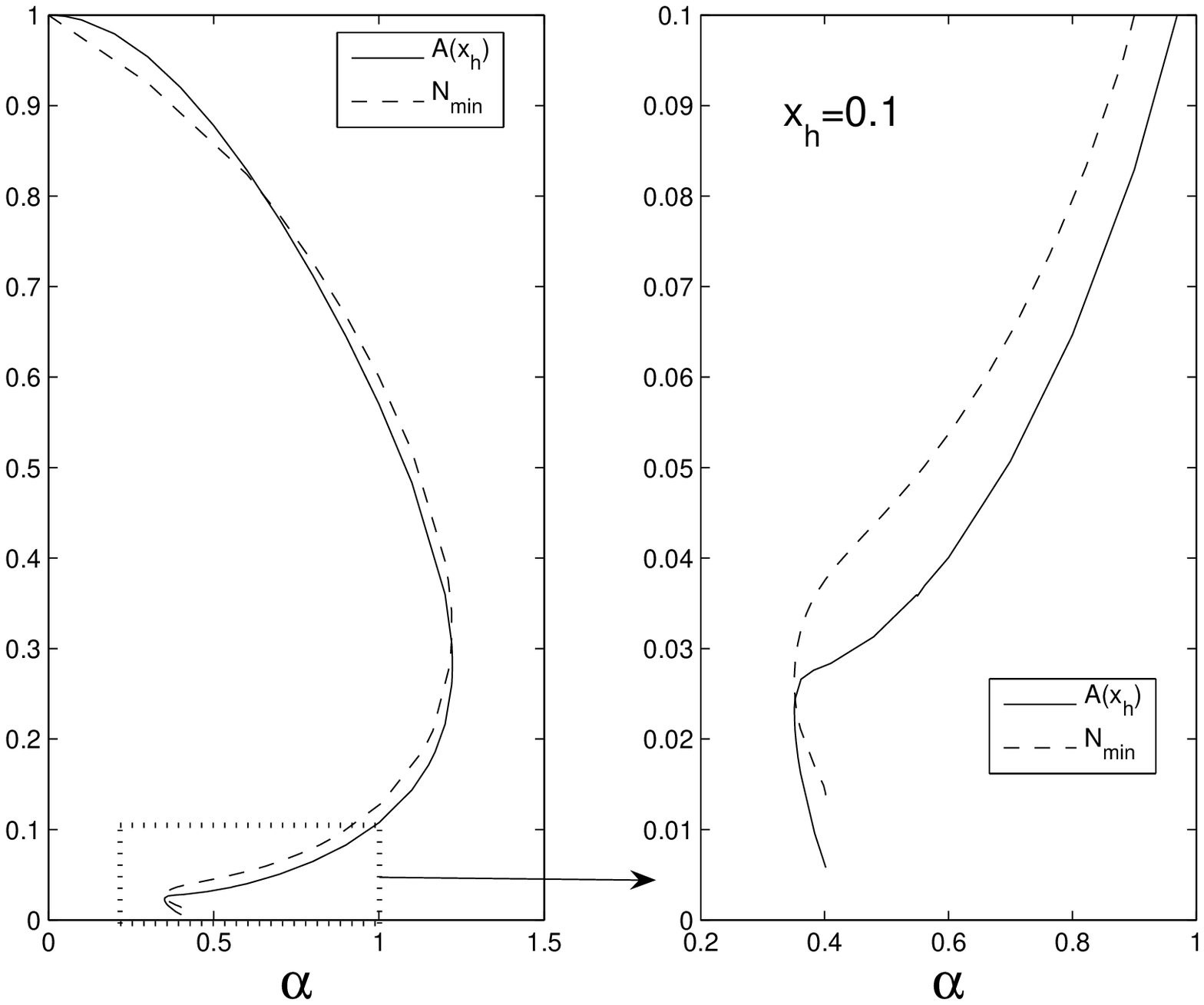}}
\caption{\label{fig1}
The value of the metric functions $A(x)$  
at the horizon,
$A(x_h)$, and the minimal value of the metric function $N(x)$ in the interval
$]x_h:\infty]$, $N_{min}$,  are shown in dependence on $\alpha$ for $x_h=0.1$.
Note that the figure on the right is an amplification of the domain
indicated by the box in the figure on the left. }
\end{figure}

\begin{figure}
\centering
\epsfysize=10cm
\mbox{\epsffile{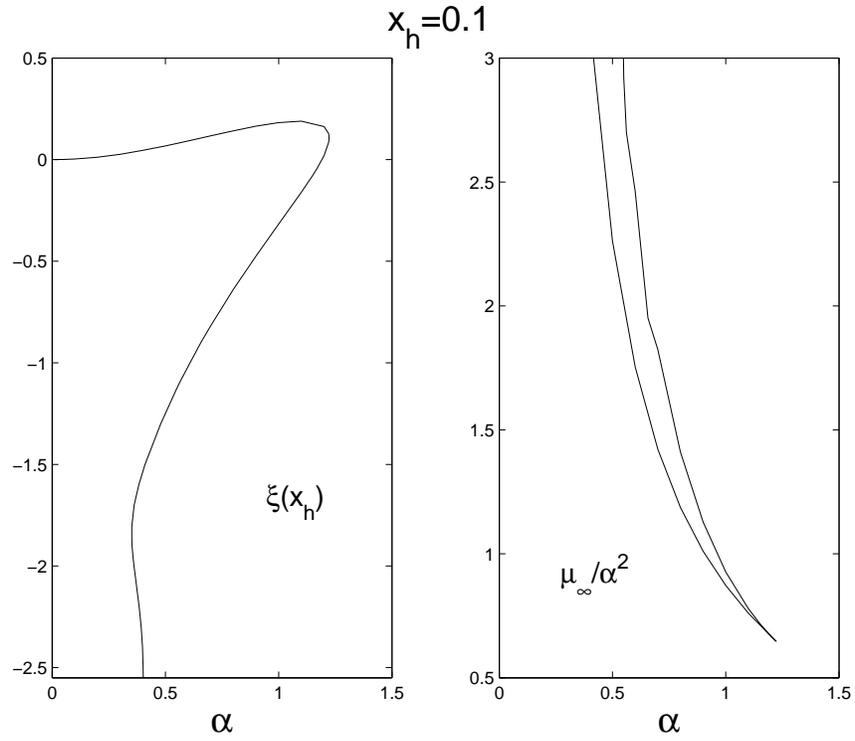}}
\caption{\label{fig2} The value of the metric function $\xi(x)$ at the horizon, $\xi(x_h)$,
is shown in dependence on $\alpha$ for $x_h=0.1$ (left). Also shown
is the dimensionless mass $\mu_{\infty}/\alpha^2$ in dependence on $\alpha$ for $x_h=0.1$ (right). }
\end{figure}

First, $x_h$ was fixed and the dependence of the solutions
on the gravitational coupling $\alpha$ was studied.
The results are shown for $x_h=0.1$ in Fig. \ref{fig1} and \ref{fig2}.
The behaviour for fixed $x_h$ is very similar to that of the
globally regular solutions \cite{volkov,bh1}. $\alpha$ was increased
from small values and the solutions exist
up to a maximal value of the gravitational coupling, $\alpha_{max}(x_h)$.
For $x_h=0.1$, we find that $\alpha_{max}(x_h=0.1)\approx 1.223$.
On this branch of solutions, the dimensionless mass $\mu_{\infty}/\alpha^2$
per unit length of the extra dimension decreases (see Fig.\ref{fig2}).
As shown in Fig. \ref{fig1},  both the value of $A(x_h)$ as well
as the minimal value of the metric function $N(x)$ in the interval $]x_h:\infty]$, $N_{min}$,
 decrease from the flat space values
$A=1$, respectively $N=1$ (see Fig. \ref{fig1}), while the value of $\xi(x_h)$
increases from $\xi=0$ (see Fig. \ref{fig2}). 
Starting from the maximal value $\alpha_{max}$,
a second branch of solutions exists up to a critical
value of $\alpha$, $\alpha_{cr}^{(1)}(x_h=0.1)\approx 0.3513$.
On this second branch $A(x_h)$ and $N_{min}$ decrease further,
while $\xi(x_h)$ now decreases, crosses zero and then becomes negative.
The mass per unit length on this branch is higher than
that on the first branch of solutions, see Fig.\ref{fig2}. 
Starting from this second branch of solutions, a third branch of solutions
exists and extends to
$\alpha_{cr}^{(2)}(x_h=0.1)\approx 0.408$. On this third branch,
$A(x_h)$ decreases further to $\approx$ zero, while $\xi(x_h)$ tends to $-\infty$.
However, $N_{min}$ stays finite. The mass on this third branch is lower than
that on the second, put differs so marginally that it cannot be seen in the plot.
 
As is clearly see, the extend of the branches in $\alpha$ decreases, which is very similar
to the globally regular case. It is likely
that an infinite number of branches exists
which extend around $\alpha\approx 0.4$.

\begin{figure}
\centering
\epsfysize=10cm
\mbox{\epsffile{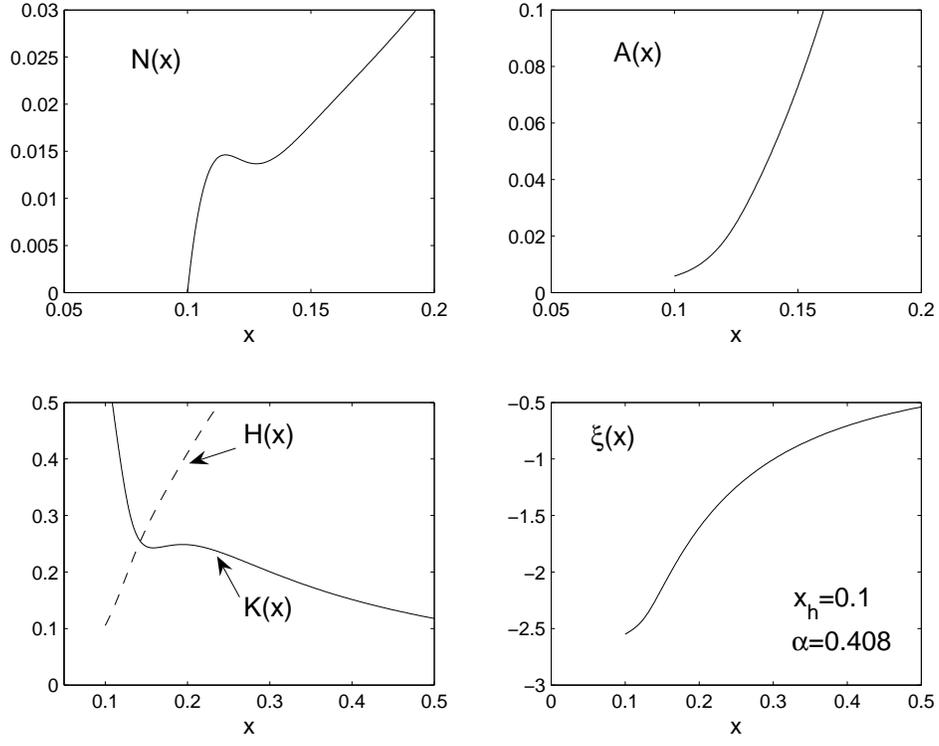}}
\caption{\label{fig3} The functions $N(x)$ (top left), $A(x)$ (top right), $K(x)$ and $H(x)$ (bottom
left) and $\xi(x)$ (bottom right) are shown close to the horizon $x_h=0.1$ for $\alpha=0.408
$.
Note that this solution belongs to the third branch of solutions. }
\end{figure}

Note that on the third branch of solutions, the function
$K(x)$ starts to develop oscillations. In Fig. \ref{fig3}, the matter and
metric functions close to the horizon $x_h=0.1$ are shown
for the solutions on the third branch for $\alpha=0.408$. $N(x)$ has developped
a local minimum. Clearly, the solution is non-trivial outside
the regular horizon and thus represents a non-abelian black string.

The oscillations of the function $K(x)$ indicate that -like
in the globally regular case- the solutions converge to the
fixed point described in \cite{volkov}. 

The fact that the solutions
exist only up to a maximal value of the gravitational
coupling was already observed in other gravitating systems containing
non-abelian gauge fields, e.g. in the 4-dimensional Einstein-Yang-Mills-Higgs
system \cite{bfm}. The first branch of solutions can be interpreted
as a branch on which the variation of $\alpha=4\pi \sqrt{G_5} v$ corresponds
to the variation of 
the gravitational coupling $G_5$, while
$v$ is kept fixed. This can be concluded from the fact that
the $\alpha=0$ limit corresponds
to the flat limit $G_5=0$. In contrast, on the further branches 
$G_5$ is kept fixed and $v$ varied. Since $v=0$ would correspond effectively
to a 4-dimensional Einstein-Yang-Mills-dilaton system, which has
Bartnik-McKinnon-type solutions \cite{bm} with 
zeros of the gauge field functions and thus different boundary conditions, 
the $\alpha=0$ limit doesn't exist for these additional branches.
The existence of a maximal value of $\alpha$ is thus related to 
the existence of a maximal possible value of the gravitational coupling $G_5$.

\begin{figure}
\centering
\epsfysize=10cm
\mbox{\epsffile{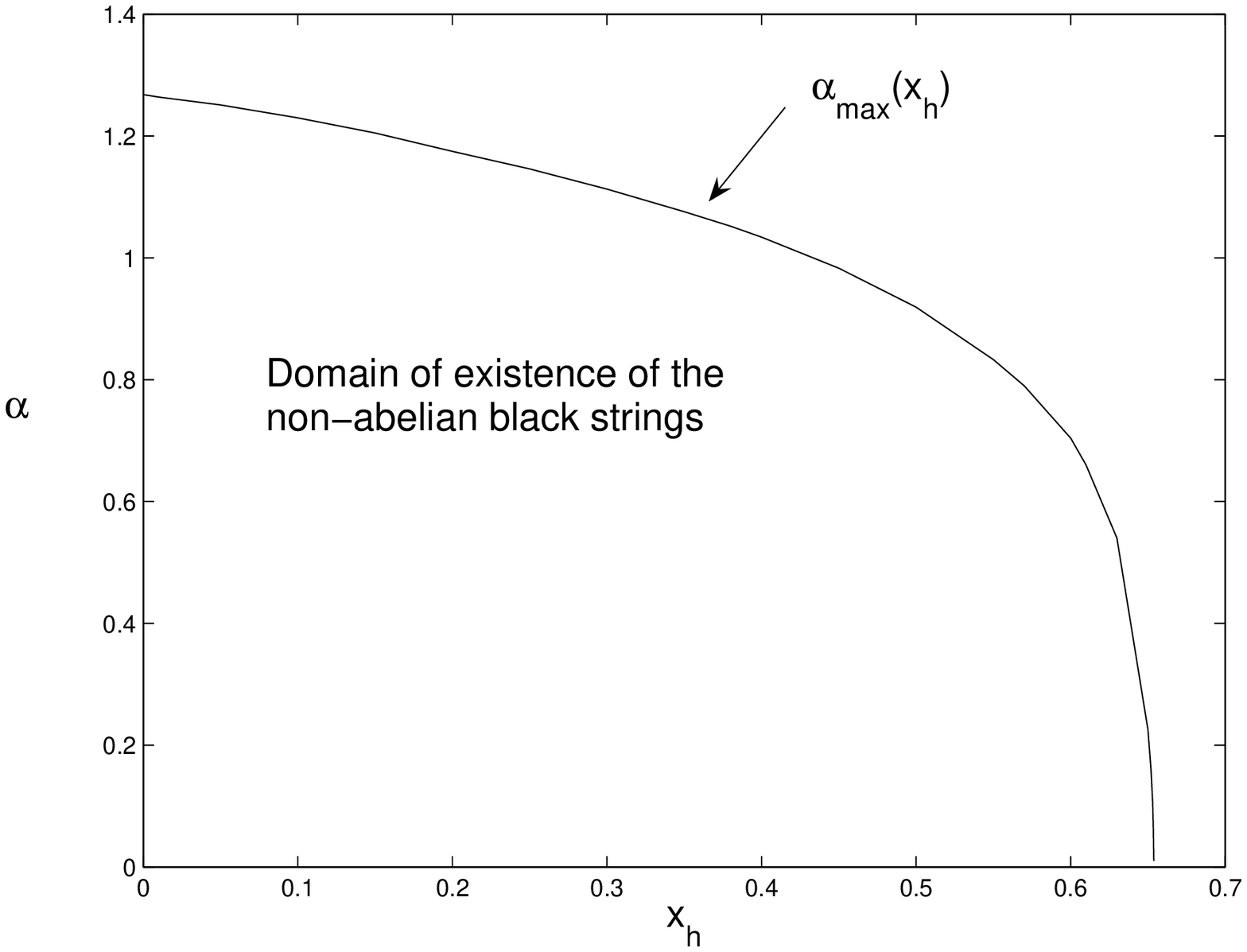}}
\caption{\label{fig4} The domain of existence of the non-abelian black strings
is shown in the $x_h$-$\alpha$-plane.  }
\end{figure}

In Fig.\ref{fig4}, the domain of existence of the non-abelian black strings is shown.
The maximal possible value $\alpha_{max}$ of the globally regular
solutions is $\alpha_{max}=1.268$ \cite{volkov,bh1}. From this value
$\alpha_{max}(x_h)$ decreases with increasing $x_h$ and finally reaches 
$\alpha_{max}(x_h)=0$ at $x_h\approx 0.654$. Non-abelian black strings
thus exist only in a limited domain of the $x_h$-$\alpha$-plane.

\section{Summary}
Black holes in more than 4 dimensions have gained renewed interest in recent years
due to theories of unification of all known forces. Since these theories
assume that the extra dimensions are compactified on a circle of the
Planck length, black hole solutions which are hyperspherically symmetric
in the full dimensions are important at early stages in the universe.
A large number of these solutions has been constructed and extensively discussed.
For smaller energies, i.e. later stages in the evolution of the universe, black hole
solutions which are spherically symmetric in 4 dimensions and extended trivially
or non-trivially into the extra dimensions are certainly of more interest.
Here, we have constructed the first example of a 5-dimensional
black string including non-abelian gauge fields. These solutions
are spherically symmetric black holes sitting in the
center of particle-like solutions in 4 dimensions, which are trivially
extended into a fifth dimension. We have shown that these solutions
exist only in a limited domain in the $x_h$-$\alpha$-plane and
that several branches of solutions (likely infinitely many) exist.
In the limit of critical coupling these solutions converge -like the globally regular
ones- to the fixed point described in \cite{volkov}.

As far as the stability of these solutions is concerned, a detailed
analysis is certainly out of the scope of this paper. However, from Morse
theory, we can assume that the second branch has one unstable mode
more than the first (since it has higher energy) and one less than the third
etc. Since for $\alpha=0$, the globally regular counterparts of the solutions
here are
effectively the solutions of a 
4-dimensional Yang-Mills-Higgs system in the BPS limit  (which
are known to be stable) it is likely that the globally regular counterparts
on the first branch of solutions are stable, while they
have $n-1$ unstable modes on the $n$th branch.
Furthermore, since for the 4-dimensional Einstein-Yang-Mills
model the number of unstable modes is equal for the globally regular and
the non-abelian black hole solutions \cite{gv}, we expect the same to be true
in the 5-dimensional analogue. This leads us to the assumption that
the non-abelian black strings are stable on the first branch and unstable
(with $n-1$ unstable modes on the $n$th branch) on the other branches.
Surely, this point has to be investigated in detail.

Further directions of investigation of the model
studied here are the construction of 4-dimensional
axially symmetric non-abelian black holes extended into one extra 
dimension or the non-trivial extension of these solutions
into the extra dimension, i.e. the construction of
non-uniform, non-abelian black strings.

\newpage

\begin{thebibliography}{99}
\bibitem{foot1}
To integrate the equations, we used the differential equation
solver COLSYS which involves a Newton-Raphson method \cite{colsys}.
\end{thebibliography}
\end{document}